\documentclass[runningheads]{llncs}
\usepackage[colorlinks=true, urlcolor=blue, linkcolor=red]{hyperref}
\usepackage[caption=false]{subfig}
\usepackage{float}
\usepackage{cite}
\usepackage[T1]{fontenc}
\usepackage{caption}
\usepackage{svg}
\usepackage{amsmath}
\usepackage{multirow}
\usepackage{graphicx}
\begin{document}

\title{Beyond Traditional Approaches: Multi-Task Network for Breast Ultrasound Diagnosis}
\titlerunning{Multi-Task Network for Breast Ultrasound Diagnosis}
%
\author{Dat T. Chung\inst{1} \and Minh-Anh Dang\inst{3} \and Mai-Anh Vu\inst{2} \and Minh T. Nguyen\inst{4}\and Thanh-Huy Nguyen\inst{5} \and Vinh Q. Dinh\inst{6}}
\authorrunning{Dat T. Chung et al.}
%
\institute{University of Technology and Education, Vietnam 
\and FPT University Ha Noi, Vietnam
\and National Cheng Kung University, Taiwan
\and Ho Chi Minh City Medicine and Pharmacy University, Vietnam
\and Taipei Medical University, Taiwan
\and Vietnamese-German University, Vietnam \\
\email{\{vinh.dq2\}@vgu.edu.vn}}
\maketitle              
\begin{abstract}
Breast Ultrasound plays a vital role in cancer diagnosis as a non-invasive approach with cost-effective. In recent years, with the development of deep learning, many CNN-based approaches have been widely researched in both tumor localization and cancer classification tasks. Even though previous single models achieved great performance in both tasks, these methods have some limitations in inference time, GPU requirement, and separate fine-tuning for each model. In this study, we aim to redesign and build end-to-end multi-task architecture to conduct both segmentation and classification. With our proposed approach, we achieved outstanding performance and time efficiency, with 79.8\% and 86.4\% in DeepLabV3+ architecture in the segmentation task.  

\end{abstract}
\section{Introduction}
Breast cancer is the leading cause of cancer deaths in women. It currently has the highest incidence rate of cancer among women in the United States; 31\% of all newly diagnosed cases of cancer in 2022 were found to be related to it. Because of its high incidence rate, early detection of breast cancer is crucial to lowering death rates and increasing available treatment options. Because breast ultrasound imaging is noninvasive, nonradioactive, and cost-effective, it is a useful screening tool. Compared to previous studies using x-rays to diagnose cancer mammograms \cite{nguyen2023context,nguyen2023towards,truong2023delving}, ultrasound diagnosis achieves more promising overall.

Deep-learning techniques have overtaken traditional ones as the preferred methods for segmenting breast ultrasound images as a result of advancements in computer technology. One of the most well-liked fully convolutional network models among them is the U-net model, which finds extensive application in the field of medical image processing. The U-net model is an end-to-end, fully convolutional network with skip layers between the synthesis and analysis paths that run pixel-by-pixel. It gained popularity due to the ability to reserve a large number of significant features using a small training dataset. Consequently, U-Net is widely used as a baseline for breast ultrasound diagnosis

In the real world, however, breast ultrasound diagnosis screening necessitates performing two tasks at the same time: segmentation and classification. In this work, we used the mechanism of multi-task learning by redesigning the framework with two distinct heads. Our proposed framework effectively uses auxiliary information from the cancer classification task to improve tumor segmentation task performance. Furthermore, we ran extensive experiments on our proposed framework under various architectures and backbones. The task weights were also carefully tested to determine how much each branch contributed to the overall loss during the training stage. Our multi-task approach achieved its robustness performance, reaching a peak at 86.4\% in the segmentation task.

\section{Related works}
\subsubsection{Medical Image Segmentation}
Artificial intelligence-based medical image segmentation has become a matter of concern in the Computer-Aided Diagnosis field. U-Net \cite{unet}, which has been widely applied in segmentation, is an autoencoder-based architecture with skip connections to incorporate feature maps from the encoder and decoder. Succeeding U-Net, several architectures have been proposed for segmentation, including U-Net++ \cite{unetpp}, DeepLabV3 \cite{deeplabv3}, and Feature Pyramid Networks (FPN) \cite{fpn} combined with various backbones as ResNet \cite{resnet}, ResNeXt \cite{resnext}, EfficientNet \cite{efficientnet}.

\subsubsection{Multi-task learning}
Multi-task learning (MTL) is an approach that aims to perform multiple related tasks simultaneously. Recently, this technique has emerged as an efficient method to boost a particular task's performance by learning valuable information from other related ones. Song et al. \cite{song} proposed a multi-task framework that performs skin lesion detection, classification, and segmentation tasks simultaneously and achieved higher scores than state-of-the-art methods, especially for segmentation tasks. Amyar et al. \cite{amyar} used a multi-task model to jointly identify COVID-19 patients and segment COVID-19 lesions from CT images, obtaining better results in both segmentation and classification. Influenced by these promising results, we implemented a multi-task model using classification tasks to boost the main segmentation task in our work.

\section{Methodology}

\begin{figure*}[t]
	\centering
	\includegraphics[width=1\linewidth]{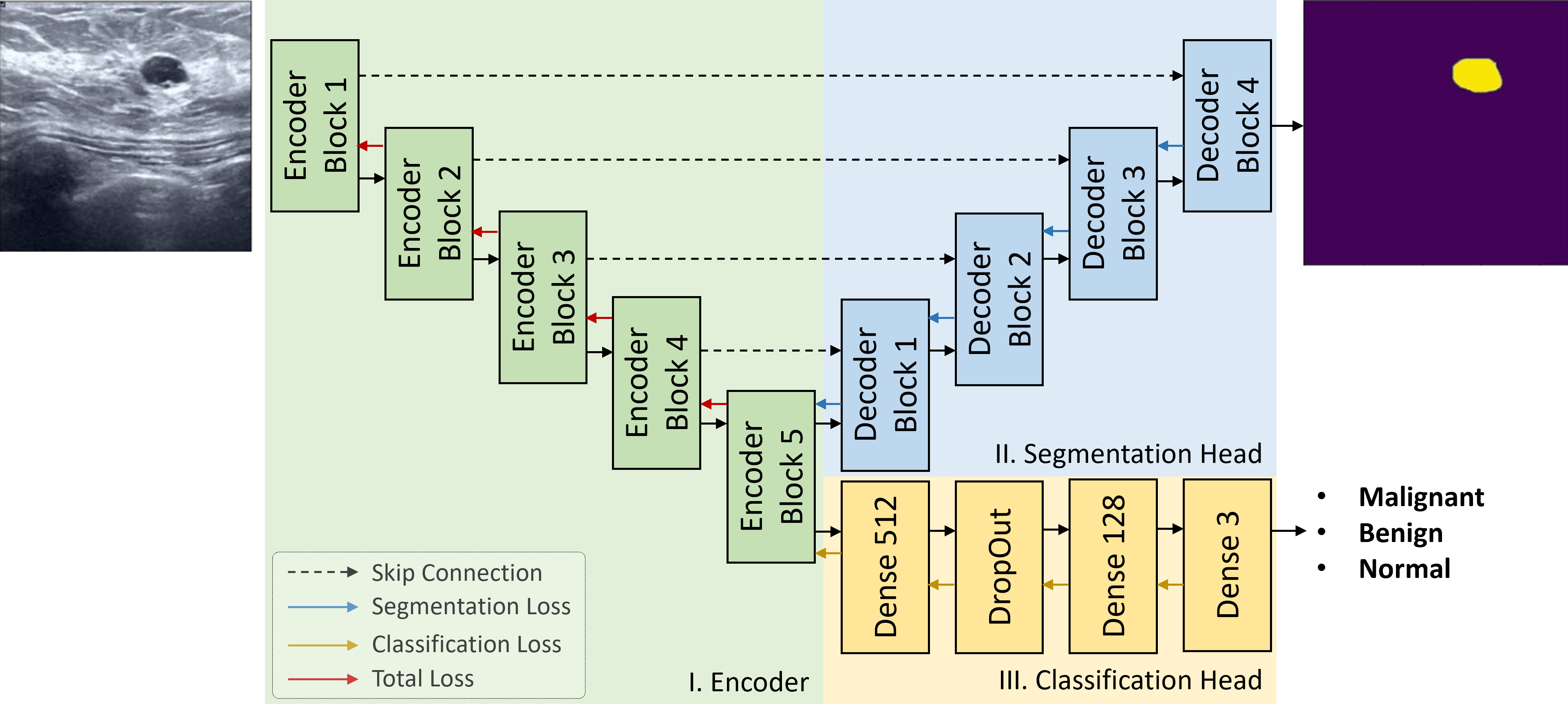}
	\caption{Visualization of our proposed architecture with multi-task learning approach. The shared encoder (green area) takes weighted sum loss from the segmentation branch (blue area) and classification branch (yellow area).}
	\label{fig: framework}
\end{figure*}

Four encoders and four decoders were chosen for their architectural diversity. The multi-task framework will consist of a shared weight backbone model with two heads handling classification and segmentation tasks. Breast ultrasound image is the input, and the outputs are the boundary of breast tumors and the cancerous attribute of breast tumors for segmentation and classification tasks, respectively.
\subsection{Multi-task framework}
We have implemented a multi-task framework that can be end-to-end trained for both segmentation and classification tasks shown in Fig~\ref{fig: framework}. Breast ultrasound images of tumors frequently show a distinct echo pattern from the surrounding tissue. To effectively capture these visual features, we use ResNet-50 \cite{resnet}, ResNeXt-50 \cite{resnext}, Wide ResNet-50 \cite{wideresnet}, and EfficientNet-B4 \cite{efficientnet}, as a shared backbone. Two separate task-specific heads are then utilized to predict breast tumor boundary and attributes.

\subsubsection{Classification head}
In particular, in the task-specific heads of classification, the extracted feature of the encoder fed into the classifier, including bias fully connected and dropout layer as illustrated in classification head of Fig~\ref{fig: framework} for prediction of either benign, malignant, or normal breast tumor.

\subsubsection{Segmentation head}
To identify the breast tumor boundaries throughout the entire image, we employ the segmentation task as a simultaneous task. The extracted features from the shared encoder will be provided through some of the most well-known decoders in the medical imaging field, including Unet \cite{unet}, Unet++ \cite{unetpp}, and DeepLabV3 \cite{deeplabv3}. Each feature will sequentially be fed into upsampling and convolution layers to recover its original size and generate the breast tumor's boundary prediction result.

\subsection{Loss composition}
\subsubsection{Classification}
A very prevalent issue that arises in medical imaging is an imbalance in the dataset. The prediction performance of the models is skewed toward the majority of samples since the number of negative samples is typically higher than the number of positive ones. Influenced by balanced cross-entropy loss and has an extra class weight parameter that can change the derivative of the loss function, Focal loss is used for the classification task. This aids in the model's attention to classes with fewer examples or those that are challenging.
\begin{equation}
L_{classification} = FocalLoss(p_{t}) = -\alpha_{t}(1 - p_{t})^{\gamma}log(p_{t})
\end{equation}

where \(p_{t}\) is the predicted probability of a ground truth class, \(\alpha_{t}\) is the balance factor for the true class, and \(\gamma\) is the hyperparameter that controls the degree of down-weighting for well-classified samples.

\subsubsection{Segmentation}
We employ Dice Loss, which is determined by averaging the dice coefficients of the background and foreground (breast tumor) classes.

\begin{equation}
L_{segmentation} = DiceLoss(y, \hat{y}) = \frac{y + \hat{y} + 2y\hat{y}}{y + \hat{y} + \epsilon}
\end{equation}

where \(y\) is the ground truth mask and \(\epsilon\) is very small to prevent loss function from division by zero. 

Finally, we argue that joint training can enhance the feature representation for each task. Thus, we define a multi-task loss as the weighted sum of two tasks:

\begin{equation}
L_{total} = \lambda L_{segmentation}(y,\hat{y}) + (1 - \lambda)L_{classification}(p_{t})
\end{equation}

where \(\lambda\) denotes the loss weight for each loss element.



\section{Experiments Results}
\subsubsection{Experimental Setting.}
We evaluated the performance of the multi-task framework using the public dataset - BUSI \cite{busi}. It consists of 780 images with an average image size of 500x500 pixels. It is categorized into three classes which are benign, malignant, and normal. In our experiments, we used an 8:2 ratio for training and test sets and resized all images to 448 × 448 pixels as the input for the model.
RTX 3090Ti 24GB is used to train all models for 50 epochs with a batch size of 8. Adam is used for optimization, with the initial learning rate set to \(10^{-4}\) and weight decay set to \(10^{-5}\). We use accuracy and F1-score metrics, which show the general effectiveness and predictive capacity of the model under imbalanced data. F1-score, Intersection over Union (IoU), and Dice Coefficient (Dice) are adopted in the evaluation of the segmentation task. 

\begin{figure*}[ht]
	\centering
        \vspace{-2mm}
        \includegraphics[width=0.6\linewidth]{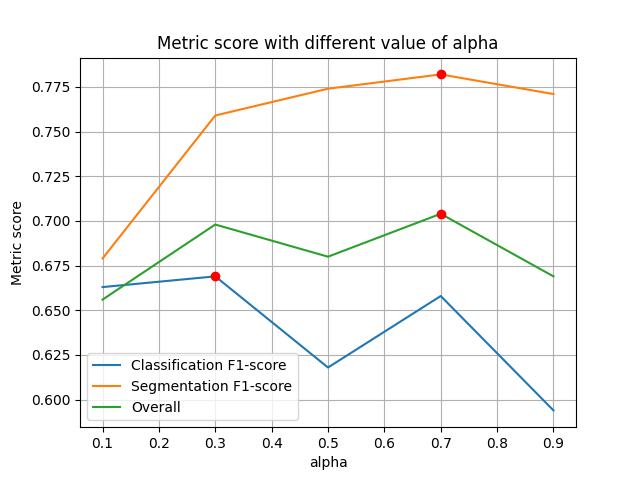}
	\caption{F1-score and Overall score of segmentation and classification task}
        \vspace{-4mm}
	\label{fig: finetune}
\end{figure*}

To determine the optimal value for hyperparameter $\lambda$ for the loss function, we perform a grid search across values ranging from 0.1 to 0.9, as depicted in \ref{fig: finetune}. The best performance for both tasks is achieved at value  $\lambda$ = 0.7.
\begin{table}[ht]
\centering
\caption{Overall performance comparison on BUSI dataset.}
\label{tab:multi-task exps}
\resizebox{9cm}{!}{%
\begin{tabular}{|c|c|cc|ccc|c|}
\hline
\multirow{2}{*}{\textbf{Architecture}} & \multirow{2}{*}{\textbf{Backbone}} & \multicolumn{2}{c|}{\textbf{Classification}} & \multicolumn{3}{c|}{\textbf{Segmentation}} & \multirow{2}{*}{\textbf{Overall}} \\ \cline{3-7}
 &
   &
  \multicolumn{1}{c|}{Accuracy} &
  F1-Score &
  \multicolumn{1}{c|}{IoU} &
  \multicolumn{1}{c|}{Dice} &
  F1-Score &
   \\ \hline
\multirow{4}{*}{Unet} &
  Resnet50 &
  \multicolumn{1}{c|}{0.682} &
  0.648 &
  \multicolumn{1}{c|}{0.682} &
  \multicolumn{1}{c|}{0.756} &
  0.756 &
  0.690 \\ \cline{2-8} 
 &
  Resnext50 &
  \multicolumn{1}{c|}{0.873} &
  0.663 &
  \multicolumn{1}{c|}{0.692} &
  \multicolumn{1}{c|}{0.758} &
  0.774 &
  0.702 \\ \cline{2-8} 
 &
  WideResnet50 &
  \multicolumn{1}{c|}{0.783} &
  0.630 &
  \multicolumn{1}{c|}{0.683} &
  \multicolumn{1}{c|}{0.763} &
  0.763 &
  0.683 \\ \cline{2-8} 
 &
  Efficientnet-b4 &
  \multicolumn{1}{c|}{\textbf{0.904}} &
  \textbf{0.723} &
  \multicolumn{1}{c|}{0.749} &
  \multicolumn{1}{c|}{\textbf{0.817}} &
  0.817 &
  0.759 \\ \hline
\multirow{4}{*}{Unet++} &
  Resnet50 &
  \multicolumn{1}{c|}{0.815} &
  0.637 &
  \multicolumn{1}{c|}{0.662} &
  \multicolumn{1}{c|}{0.746} &
  0.746 &
  0.678 \\ \cline{2-8} 
 &
  Resnext50 &
  \multicolumn{1}{c|}{0.822} &
  0.651 &
  \multicolumn{1}{c|}{0.760} &
  \multicolumn{1}{c|}{0.771} &
  0.838 &
  0.720 \\ \cline{2-8} 
 &
  WideResnet50 &
  \multicolumn{1}{c|}{0.790} &
   0.664 &
  \multicolumn{1}{c|}{0.680} &
  \multicolumn{1}{c|}{0.761} &
  0.761 &
  0.699 \\ \cline{2-8} 
 &
  Efficientnet-b4 &
  \multicolumn{1}{c|}{0.847} &
  0.650 &
  \multicolumn{1}{c|}{0.769} &
  \multicolumn{1}{c|}{0.795} &
  0.837 &
  0.725 \\ \hline
\multirow{4}{*}{FPN} &
  Resnet50 &
  \multicolumn{1}{c|}{0.688} &
  0.627 &
  \multicolumn{1}{c|}{0.606} &
  \multicolumn{1}{c|}{0.601} &
  0.668 &
  0.626 \\ \cline{2-8} 
 &
  Resnext50 &
  \multicolumn{1}{c|}{0.745} &
  0.652 &
  \multicolumn{1}{c|}{0.714} &
  \multicolumn{1}{c|}{0.752} &
  0.794 &
  0.703 \\ \cline{2-8} 
 &
  WideResnet50 &
  \multicolumn{1}{c|}{0.758} &
  0.547 &
  \multicolumn{1}{c|}{0.710} &
  \multicolumn{1}{c|}{0.749} &
  0.790 &
  0.648 \\ \cline{2-8} 
 &
  Efficientnet-b4 &
  \multicolumn{1}{c|}{0.796} &
  0.680 &
  \multicolumn{1}{c|}{0.741} &
  \multicolumn{1}{c|}{0.776} &
  0.817 &
  0.729 \\ \hline
\multirow{4}{*}{DeepLabV3+} &
  Resnet50 &
  \multicolumn{1}{c|}{0.809} &
  0.676 &
  \multicolumn{1}{c|}{0.715} &
  \multicolumn{1}{c|}{0.769} &
  0.794 &
  0.718 \\ \cline{2-8} 
 &
  Resnext50 &
  \multicolumn{1}{c|}{0.726} &
  0.651 &
  \multicolumn{1}{c|}{0.745} &
  \multicolumn{1}{c|}{0.781} &
  0.823 &
  0.717 \\ \cline{2-8} 
 &
  WideResnet50 &
  \multicolumn{1}{c|}{0.809} &
  0.610 &
  \multicolumn{1}{c|}{0.722} &
  \multicolumn{1}{c|}{0.797} &
  0.797 &
  0.691 \\ \cline{2-8} 
 &
  Efficientnet-b4 &
  \multicolumn{1}{c|}{0.892} &
  0.694 &
  \multicolumn{1}{c|}{\textbf{0.798}} &
  \multicolumn{1}{c|}{0.813} &
  \textbf{0.864} &
  \textbf{0.760} \\ \hline
\end{tabular}%
}
\end{table}


\subsubsection{Overall Results}


For the classification task, EfficientNet-B4 consistently outperforms nearly all other backbone models across different architectures. It takes the top position with DeeplabV3+, FPN, and Unet while taking the second position with Unet++. Remarkably, EfficientNet-B4 maintains competitive performance despite utilizing fewer parameters when employed as the segmentation head. Resnet50 and Resnext50 alternate between the second and third positions. 

Regarding segmentation results, EfficientNet-B4 stands out for its consistent and competitive performance across segmentation models, demonstrating its efficacy. 
EfficientNet-B4 consistently achieves top rankings for IoU, Dice, and F1-score, showcasing high performance across many architectures like DeeplabV3+, FPN, Unet, and Unet++.

In summary, the combination of EfficientNet-B4 and Unet demonstrates the best performance in both methods, encompassing most of the classification metrics and Dice for segmentation. Meanwhile, the pairing of EfficientNet-B4 with DeepLabV3++ achieves the highest IoU and F1-score for segmentation, with a Dice result closely approaching the top rank. In conclusion, EfficientNet-B4 exhibits effectiveness across all segmentation methods. 

\subsubsection{Visualization}

\begin{figure*}[htb]
    \vspace{-10mm}
    \centering
    \includegraphics[width=0.7\linewidth]{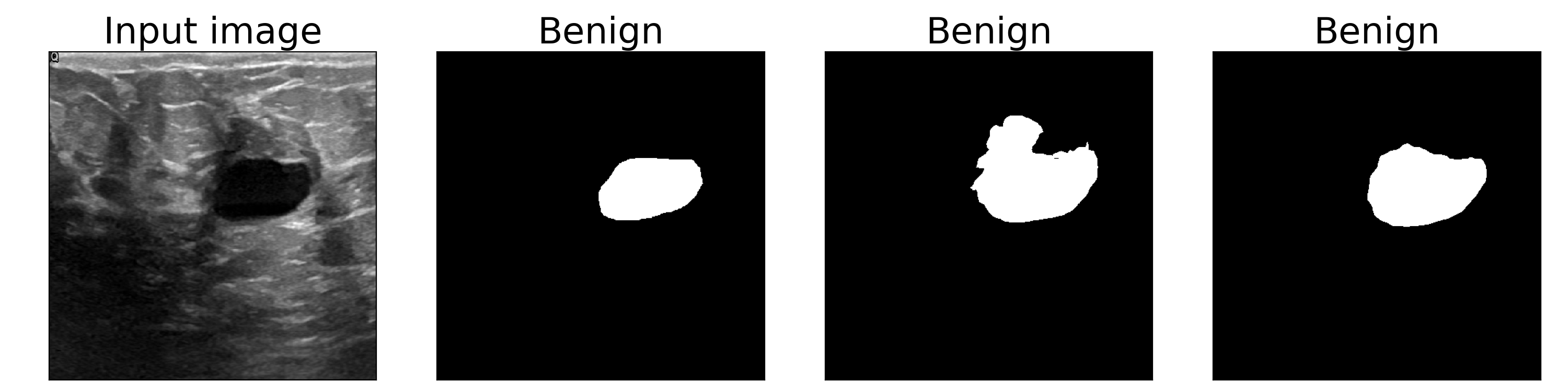} \\
    \includegraphics[width=0.7\linewidth]{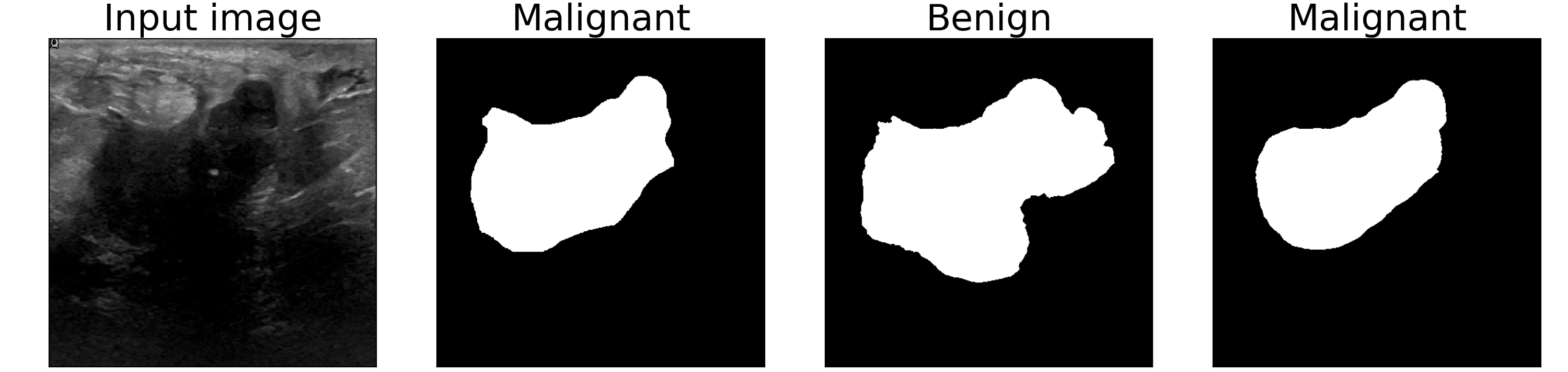} \\
    \includegraphics[width=0.7\linewidth]{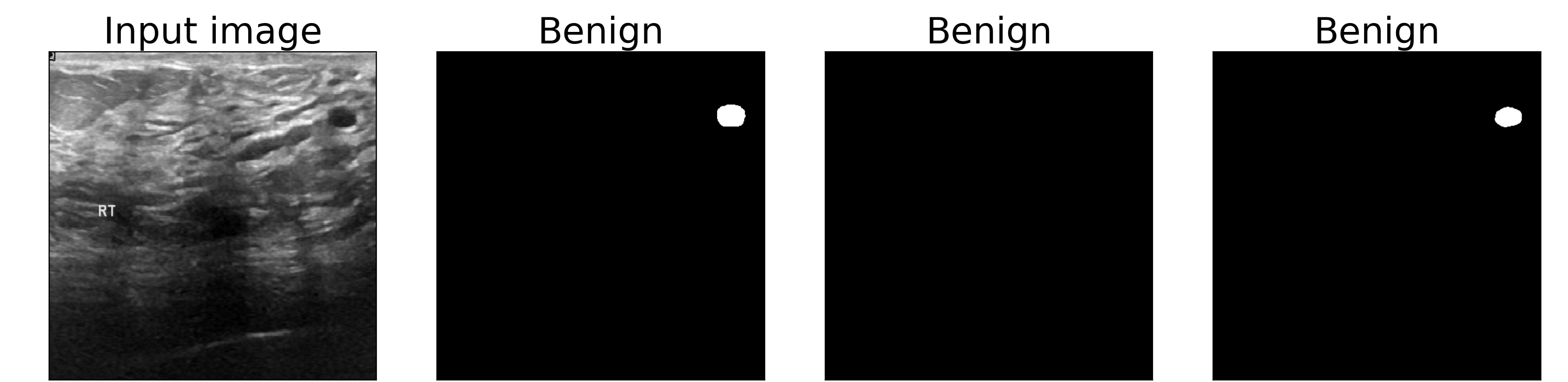}
    \caption{Visualization of prediction results of the naive approach and our multi-task framework. (a) Breast ultrasound image; (b) Ground truth masks; (c) Naive segmentation approach output; (d) Our multi-task framework output.}
    \label{fig: visualize}
    \vspace{-4mm}
\end{figure*}
Qualitative predictions of the best-performing model are displayed in Fig.~\ref{fig: visualize}. The segmentation predictions of the multi-task framework look strong, with small gaps from the ground truth. In particular, the middle visualization shows our proposed method, in contrast to the naive approach, can identify the position of a very small tumor with a boundary quite well. Additionally, in the bottom illustration, the tumor attribute is also precisely predicted, and the segmentation result is extremely close to the ground truth. This demonstrates how information from segmentation and classification tasks complements each other and helps to enhance boundary and tumor attribute prediction.

\section{Conclusion}
In this work, we used Breast Ultrasound Images to examine the efficacy of multi-task learning. Our network with a shared backbone and two heads was proposed to effectively conduct two tasks.
Furthermore, we empirically demonstrate the efficacy of the hyperparameter $\lambda$ weight in the loss function to determine the potency of which each task can effectively contribute to the others. We intend to extend this straightforward methodology to tackle the gradient projection among different goal functions and the dependency problem in multi-task learning.

\bibliography{reference}
\bibliographystyle{ieeetr} 
\end{document}